  \documentclass[aps,graphicx,twocolumn,prb,tightenlines]{revtex4}
\usepackage[dvips]{graphicx}
\usepackage{psfrag,subfigure,color}
\newcommand{\ket}[1]{\left |#1\right \rangle}
\newcommand{\bra}[1]{\left \langle #1\right|}

\newcommand{\lbulk}{L_{\rm bulk}}
\begin{document} 
%
%
\title{Nonchiral Edge  States at the  Chiral Metal Insulator  Transition in  Disordered Quantum Hall Wires}  
\author{Alexander Struck$^{(1)}$,  Bernhard 
 Kramer$^{(1)}$, Tomi  Ohtsuki$^{(1,2)}$,  Stefan Kettemann$^{(1)}$ }  
\affiliation{ $^{1}$I. Institut f\"ur 
  Theoretische Physik, Universit\" at Hamburg, Jungiusstra\ss{}e 9, 20355 
  Hamburg, Germany\\ 
 $^{2}$Department of Physics, Sophia 
  University, Kioi-Choi 7-1, Chiyoda-Ku, Tokyo 102-8554, Japan} 
\begin{abstract} 
 The  quantum phase diagram  of 
disordered  wires in a 
  strong magnetic field is studied as function of wire width and energy. 
  The 2--terminal conductance shows     zero temperature discontinuous transitions 
  between {\it exactly} integer
  plateau values and zero.
 In the vicinity of 
   this transition, the chiral metal insulator transition (CMIT),  
  states  are identified  
 which are  superpositions of 
 edge states with opposite chirality. The bulk contribution of such 
  states is found  to decrease with increasing wire width. 
    Based on exact diagonalisation results for the 
     eigenstates and their participation ratios,  we conclude that  
  these states are characteristic for the CMIT, have the 
   appearance of  nonchiral edges states, 
  and  are thereby   distinguishable 
 from  other states in the quantum Hall wire, 
 namely, 
 extended edge states,  two-dimensionally (2D) localized,  quasi-1-D localized, 
 and  2D critical states. 
\end{abstract} 
\maketitle 
\vskip2pc
 
\section{Introduction}
 
 Recently, there has been renewed  interest in 
 quantum Hall bars of finite width, 
 where the interplay between  localized
states in the bulk of the 2-dimensional electron system (2DES) and 
edge states with energies lifted by the confinement potential above the
energies of  centers of  bulk Landau bands, $E_{n0}$,  \cite{halperin} 
 results  in  the quantization of the Hall conductance. 
The study of
mesoscopically narrow quantum Hall bars, \cite{haug}
 revealed  new types of  conductance
fluctuations, \cite{timp,ando94} edge state mixing,
 \cite{edgemixing,shkledge,ohtsuki,ando,mani}  the  breakdown of the quantum Hall
effect, \cite{breakdown} and  the quenching of the Hall effect due to classical
commensurability effects.  \cite{roukes}
In the presence of white noise disorder
the edge states do mix with the bulk states when the Fermi energy is moved
into the center of a Landau band. It had been suggested that this might result
in localization of edge states.  \cite{ohtsuki,ando1990,mani}
 Recently, it has been  shown   that
  at zero temperature the
two-terminal conductance  of a quantum wire in a
magnetic field exhibits for uncorrelated disorder and hard wall confinement
discontinuous transitions between integer plateau values and zero. \cite{jetp}
 These transitions have been argued  
 to be due to sharp localization transitions  of  chiral edge states, 
 where   the localization length of the edge states
 jumps  from  {\it exponentially large}
 to finite
  values, driven by the dimensional crossover of localized bulk states,
   and are accordingly called chiral metal insulator transitions (CMIT).

 In this article, we will study the nature of this transition in 
 more detail, and in particular find that at this transition there
 exists a new type of state, 
 with properties distinguishable from both localized and extended bulk states, 
 and extended edge states. 
 This new state is a superposition  of
 edge states with opposite chirality. Since it is still located mainly 
  close to  the edges, we will call  this state {\it nonchiral edge state}.  

 The article is organised in the following way. 
 In the next section, we present transfer matrix  calculations
  of the  quantum 
 phase diagram of a quantum Hall bar with uncorrelated disorder, 
 being characterised by the two--terminal conductance $G$ as function of 
 energy $E$ and width $w$ of the wire. 
 Sharp jumps in the conductance
  from integer values to zero are found as function of energy.
   These  CMIT's,  are seen 
  to  become more pronounced with increasing  wire widths $w$.

 In the third chapter we will study with exact diagonalization 
  the eigenstates of a disordered quantum Hall wire. 
 We will classify these states into five classes, 
 the edge states, the 2D localized states, the quasi-1-D localized states, 
 2D extended states, and the new nonchiral edge states at the 
 chiral metal insulator transition. 
 These states are characterised by their specific participation ratio as function of energy and wire width $w$, 
 their distribution of coefficients in an expansion in eigenstates of the clean 2DES, and 
 the spatial distribution of the 
 eigenfunction amplitudes. 
  This allows us to identify the state at the transition as 
 a superposition  of edge states of opposite chirality. 
 
 



 The final chapter contains our conclusions, and a discussion on 
 how the CMIT  could be observed experimentally.

\section{The Quantum Phase Diagram of the CMIT}
 \begin{figure*}[htbp]
\begin{center} 
\vspace{1cm} 
\includegraphics[width=.8 \textwidth]{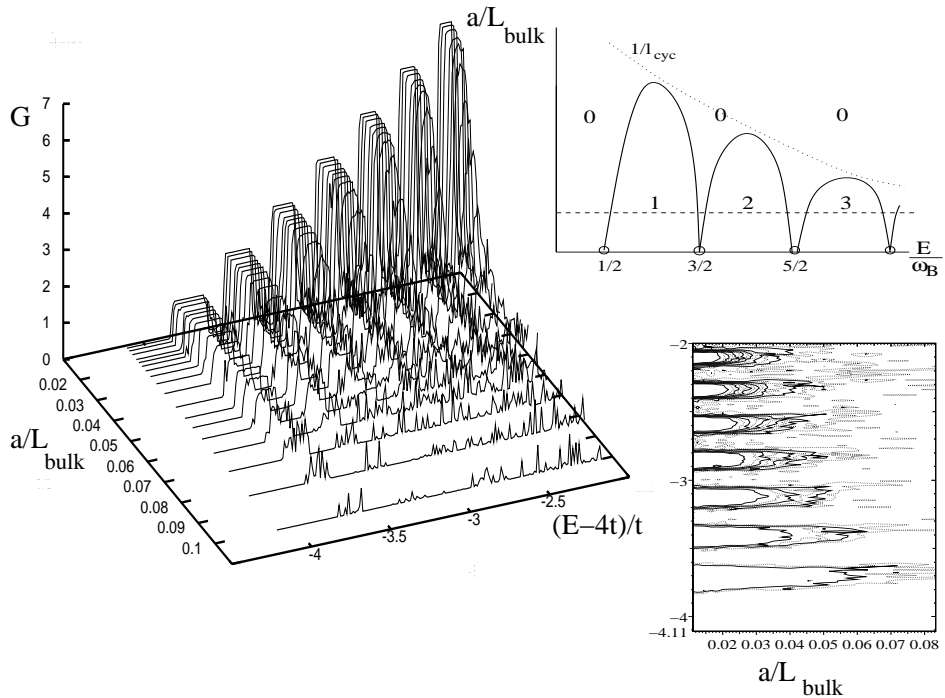} 
\caption[]{ The conductance as function of energy 
 for increasing values of the  width $\lbulk$
  (left), and in a contour plot as function of 
   energy and $\lbulk$ (lower right), as compared with the schematic phase diagram (upper right). Finite integer values of the 
   conductance correspond to the number of extended edge states. The disorder is uniformly distributed in an interval of width 
   $W =0.8 t$. There are $x=0.025$ magnetic flux quanta per elementary cell of area $a^2$.  }
 \label{phasenew} 
\end{center} 
\end{figure*} 
Using the transfer matrix method,  \cite{transfer} we have calculated the  2-terminal conductance\cite{pmr92} $G$ 
 as function of energy $E$ in a tight binding model with band width $ 8 t$, where $t$  is the hopping amplitude,  of a
disordered quantum wire in  a perpendicular magnetic field  (Fig. \ref{phasenew}),
 with hard wall boundary conditions at $y=\pm \lbulk/2$ and finite length $L= 2000 a$.\cite{jetp} Here we have assumed a square lattice with lattice spacing $a$.  The disorder potential 
is uniformly distributed in  an interval $[- W/2, W/2 ]$.  
  These results are  summarised in the phase diagram (Fig. \ref{phasenew}), 
 where the value of  $G$, in units of $e^2/h$, 
 is given  as function of bulk width $\lbulk$ and energy $E$ in units of $\hbar \omega_B$, for a disorder strength $W = 0.8 t$. 
 As expected,   $G= m$,
 where $m$ is the number of extended edge states between 
the   Landau bands. 
 Close to the middle of the Landau bands, however,  the   conductance plateaus
  collapse abruptly to $G=0$.
 
\begin{figure*}[htbp]
\begin{center}
\vspace{1cm} 
\includegraphics[scale=.6, angle=-90]{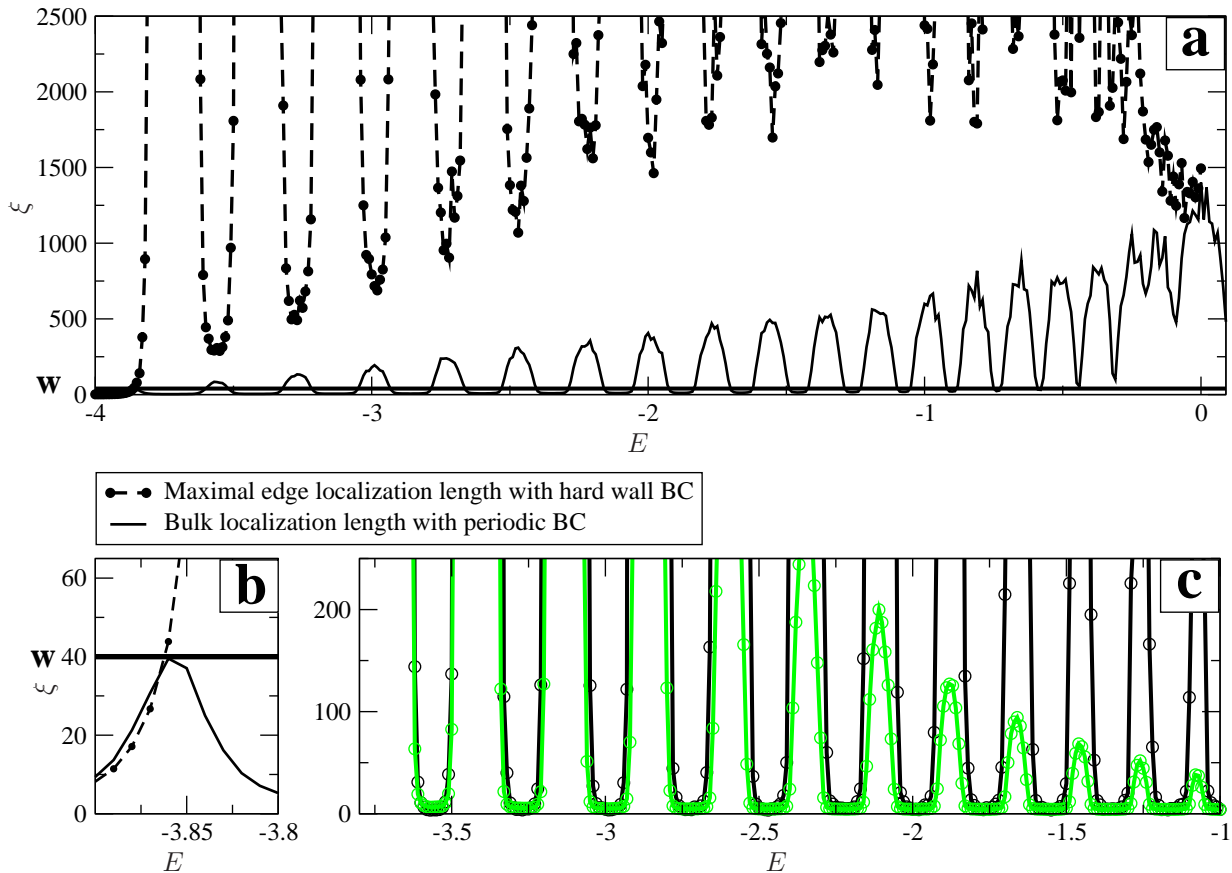} 
\caption[]{(Color online) (a) The localization length for a disordered wire
  calculated with the transfer matrix method with periodic boundary conditions
  (full line) and with hard wall boundary conditions (dashed line) for uniformly distributed uncorrelated disorder in an interval of 
  width $W=0.8 t$.  
  The straight line indicates the bulk width $w=40 a$. There are $x=0.025$ magnetic flux quanta per elementary cell of area 
  $a^2$. (b) Enlargement of the low-energy region of (a). Edge and bulk localization length coincide as long as no edge state is 
  present. (c) The functions $\xi_{\rm edge}/\xi (E)$ (black) and $e^{2w/\xi}$ (green). } 
 \label{lcedge} 
\end{center} 
\end{figure*} 

  When the  wires are so narrow, that the edge states cannot form, as it is the case when the width is smaller than the 
  cyclotron length,  or when edge states of opposite chirality are mixed by backscattering, then all the   states become  
  localized and 
 the conductance is vanishing with only small mesoscopic fluctuations  due to the 
 finite length $L$ of the wire.   
 Previously, it has been  pointed out that,  
 when the bulk localization length $\xi$ is smaller than the physical wire width $w$,  backscattering between edges is exponentially 
suppressed. As a result, the localization length of edge states 
 increases strongly.

 The overlap of opposite edge states is known to decrease
  exponentially  with increasing
 wire width $w$ \cite{shkledge}.  Thus, the backscattering rate 
between edges, being proportional to the 
 square of the overlap integral  is  $1/\tau  \sim \exp (-2 w/\xi )$.
  Since the edge states are one-dimensional, their localisation length due to
  the back scattering is given by $\xi_{edge} = 2 v_F \tau$, with Fermi velocity $v_F$. 
  On the other hand, when the bulk localization length $\xi$  becomes equal to
  the wire width, one expects that the edge states
 become mixed with the bulk states, and  localized with a length proportional
  to the 
  bulk localization length $\xi$. Therefore, we conjectured the edge localization length to behave like \cite{jetp}    
  \begin{equation}
  \label{edge}
  \xi_{\rm edge} = \xi \exp (2 w/\xi),
  \end{equation}  

 In Fig.  \ref{lcedge}a,  the localization length 
 is plotted as function of energy,
  as obtained with the transfer matrix method 
in a tight binding model of a
disordered quantum wire in  a perpendicular magnetic field
  with hard wall boundary conditions, 
   dashed line. Since the edge states are the most extended
    states in the wire, this localisation length 
     can be identified with the edge localisation length $\xi_{edge}$. 
   
Using the transfer matrix method,  \cite{transfer} we have also calculated the
localization length $\xi$ as function of energy $E$ for a 
disordered quantum wire  with idendical properties, but  with periodic
boundary conditions, Fig.  \ref{lcedge}a, solid curve. Since there are   no edge states 
 this  bulk localization length is small in the tails of Landau bands, and has 
maxima, which are  seen to increase linearly with  $n$.

   Indeed the behavior of the edge state localisation length 
   follows qualitatively the behaviour suggested by  Eq. (\ref{edge}). The edge localization length does increase sharply, whenever
    the bulk localisation length  becomes smaller than  the 
    wire width $w$ (full straight line). Note that the minima  in the
 middle of the Landau bands do increase linearly with the 
 Landau band number $n$. In Fig. \ref{lcedge}c, we have explicitly 
 plotted $\xi_{edge}/\xi$ and $ \exp ( 2w/\xi )$, using the numerically
  calculated values for $\xi_{edge}$ and $\xi$, as function of $E$.
  We find that both functions coincide for all energies above the 
    lowest Landau band and 
     for $\l_{cyc} < w$, so that edge states exist in the tails of 
   the Landau bands. 
 
 An abrupt decrease of the inverse localisation length 
  has been found before for energies in  the upper tail of the 
  lowest Landau level and the tails of the second Landau level 
   in Ref. 8.
   In agreement with our above results, it has been found there, that 
    the inverse localisation length decays exponentially in the tails, 
     like $1/\xi_{edge} \sim \exp ( - \beta (E)  w)$. The fitted values of $\beta$ have been found there to depend weakly
       on energy, whereas we identified it directly with the 
         energy dependent inverse bulk localisation length 
          $1/\xi(E)$.

From these results we can  conclude that  the energy at which   edges states backscatter and become localized, is, 
 given by the condition that   the bulk localization length is on the order of  the wire width, $\xi(E_{m,p}) = w$. 
 At this energy,  $m$ edge states mix and
transitions from  extended edge states to insulating states  occur. This causes sharp 
jumps of $G$ from finite integer  to vanishingly small values, as seen  in Fig. \ref{phasenew}. This can be explained by the  exponential decrease of the edge state localization length, Fig. \ref{lcedge}.  We note that  $m=n$
when the energy is above the $n$-th Landau band , whereas $m= n-1$, if it is below.

 A more detailed  understanding of this 
  drastic  behaviour of the conductance can be obtained  by considering 
   the dimensional crossover of the 
   bulk localization length in disordered wires. \cite{crossover,jetp}
In a 2DES with broken time reversal symmetry,  
  scaling theory \cite{ab,weg,hikami,elk} and
numerical scaling studies  \cite{mk81,transfer} find that the bulk localization length $\xi$ is independent of the wire width, 
$\xi_{\rm 2D}  = l_0\, \exp (\pi^2 g^2)$.
 Here,  
 $g$,  is the 2D conductance parameter per spin channel.  
 $l_0$ is the short distance cutoff, 
 the elastic mean free path $l = 2 g(B=0)/k_{{\rm F}}$ ($k_{\rm F}$ Fermi
wave number) at weak  magnetic fields, $b\equiv\omega_B \tau < 1$.
 For stronger magnetic fields, $b>1$,  the short length scale  $l_0$ becomes   the cyclotron length $l_{\rm cyc}$.
 The conductance parameter    $g$ exhibits Shubnikov-de-Haas oscillations as function
of  magnetic field for $b>1$. Maxima occur when the Fermi energy is in
the center of  Landau bands.
 The localization length in  tails of  Landau bands, where $g \ll 1$ is very small,
 is of the order of the cyclotron length
 $l_{\rm cyc} = v_F/\omega_B = \sqrt{2 n+ 1} l_B$. 
   It increases towards the centers of the Landau
 bands, $E_{n0} = \hbar \omega_B (n +1/2)$ ($n=0,1,2,...$), with $\omega_B = e
 B/m^{*}$ the cyclotron frequency ($e$ elementary charge, $m^{*}$ effective
 mass),  $v_F$ the Fermi velocity, 
 and $l_B^2 = \hbar/e B$ defines the magnetic length.
   In an {\em infinite} 2DES in 
 perpendicular magnetic field, the localization length at energy $E$ diverges
 as $ \xi \sim |E -E_{n0}|^{-\nu}$.  The critical exponent $\nu$ is known from  numerical
  finite size scaling studies  for the lowest
 two Landau bands, $n=0,1$, to be $\nu = 2.33\pm0.04$ for spin-split Landau
 levels,  \cite{hucke,huckerev} in agreement with  analytical  \cite{raikh} and
 experimental studies  \cite{scalingexp}. In a
 {\em finite} 2DES, a region of  state exists in the centers of
  disorder broadened Landau-bands, which cover  the whole system
 of  size $L$. The width of these regions is given by $\Delta E = (l_{\rm
   cyc}/L)^{1/\nu} \Gamma$, where   $\Gamma = \hbar ( 2
 \omega_B/\pi\tau)^{1/2}$ is the band width,
 with elastic scattering time $\tau$. 

However, the  2D localization length is seen  to
increase strongly from  band tails to  band centers, even 
 when  the wire  width $w$  is so narrow, that it is far from the critical point at  $w \rightarrow \infty$. 
One can estimate the noncritical localization length for 
 uncorrelated impurities, by inserting $g$, as obtained within self
consistent Born approximation.  \cite{ando}  Its  maxima 
 are  $g(E = E_{n 0}) =  (2  n + 1)/\pi = g_n
$.  Thus, $ \xi_{\rm 2D} (E_{n 0}) = l_{\rm cyc} \exp{( \pi^2 g_n^2 )} $ are macroscopically
large in  centers of higher Landau bands, $n > 1 $.
 \cite{huckerev,levitation}
  When the width of the system $w$ is
smaller than  $\xi_{\rm 2D}$,  electrons in  centers of  Landau
bands can diffuse between the edges of the system. In long wires, however,  the electrons are localized due to quantum interference  along the wire with 
 a localization length that is found to depend  linearly on $g$ and $w$,  \cite{ef,larkin,dorokhov,jetp}
\begin{equation} \label{1db} 
\xi_{\rm 1D} = 2 g (B)  w.   
\end{equation}  
The conductance  per spin channel, $g(b) =
\sigma_{xx}(B)/\sigma_0$, is given by the Drude formula $ g(b) =
g_0/(1+b^2)$, ($g_0 = E \tau/\hbar$, $b=\omega_B \tau$) for weak magnetic field, $b
<1$.  For $b>1$, when the cyclotron length $ l_{\rm cyc}$ is smaller than the
mean free path $l$, disregarding the overlap between  Landau bands, $g$ is
obtained in SCBA,  \cite{ando}
$
g(B)  = (1/\pi)  (2 n+1) ( 1 - (E_{\rm F}- 
 E_n)^2/\Gamma^2 ),  
$
for $ \mid E - E_n \mid < \Gamma$.  One obtains the localization length for $b>1$ and
$\mid \epsilon/b -n -1/2 \mid < 1$ by inserting $g$.
 It oscillates between maximal values in
 centers of  Landau bands, and minimal values in  band tails.
  For $n>1$, one finds in  band centers,
\begin{equation} \label{middle} 
\xi_n =  \frac{2}{\pi} \left( 2 n + 1 \right) w   
\left[ 1- \frac{\ln 
\sqrt{ 1+ \left(w/l_{\rm cyc} \right)^2}}{(n+1/2)^2} \right]^{1/2}.    
\end{equation}
 Thus, the localization length in the center of Landau bands
  is found  to increase linearly  with Landau band index $n$. 
   This is exactly the behaviour observed above  in the numerical results, Fig. \ref{lcedge}.

 While it is reasonable to conclude that the edge states do mix with the bulk
 states at the energy where the bulk localization length is equal to the wire width,
 and the electrons diffuse freely from edge to edge but are localized along the wire, 
  the question arises, how exactly this transition from extended edge states
 to localized states  does occur.  One can gain some 
  further insight by  connecting the two ends of the wire together  to form   an annulus. 
   Piercing    magnetic flux through   the annulus affects only  states whose localization length is larger 
   than the circumference of the annulus.
    Guiding centers of those states which extend around the
annulus do shift in position and energy  \cite{halperin} with a change in  magnetic
flux. As shown above,  in the middle of the Landau band, the  electrons can
diffuse freely from edge to edge, but are localized along the annulus
 with $\xi > w$.  When adiabatically changing the magnetic flux, the energy of an edge state changes continously. However, it  cannot enter the band of
localized states, so that  at the
energy $ E_{m}$, with $\xi (E_m) = w$,
 the edge state must be transfered to the opposite edge. There it moves up in energy when the magnetic flux is increased 
 further. \cite{halperin}
 
 In the following, we study the states at this transition in detail in order to find out, whether the edge states become localized mainly by 
 mixing completely with the bulk states, or  rather 
  the transition from 
  extended chiral edge states to localized states  occurs 
   due to the nonlocal coherent superposition  of  edge states with opposite chirality,  located at  opposite edges. 

\section{Exact diagonalization}
In this chapter, we study  the localization properties of electrons in quasi one-dimensional wires in the presence of disorder and a strong magnetic field by means of exact diagonalization.
\subsection{The model}
\label{q1Dmodel}
The Hamiltonian of the quasi-1D-wire in the presence of a disorder potential
$V_{\rm dis}$ and a confinement potential $V_{\rm conf}$, is given by  
\begin{equation}
H=\frac{1}{2m^\ast}\left({\bf p}- e {\bf A}\right)^2+ V_{\rm dis}({\bf r})+ V_{\rm conf}({\bf r})\; ,
\label{Hamiltonian_diag}
\end{equation}
 where $e>0$ is the elementary charge and $m^\ast$ the effective electron mass.

The disorder potential is modelled as
\begin{equation}
V_{\rm dis}({\bf r})=\sum\limits_{i=1}^{N_{\rm imp}}V_i \delta({\bf r}-{\bf r}_i),
\end{equation}
where $N_{\rm imp}$ is the number of impurities with uniformly distributed amplitude  $V_i\in [-V_0,V_0]$.
  ${\bf r}_i$ is  the random position of
the impurity.

\begin{figure}[htbp]
{\hspace*{-0.8cm}\includegraphics[scale=0.8]{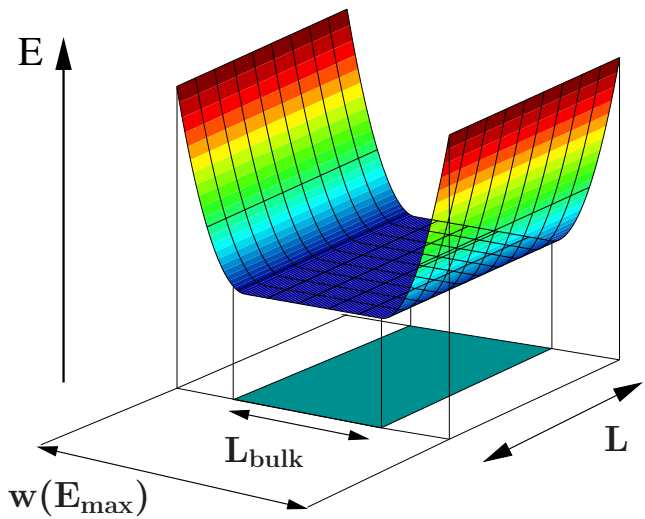}}
\caption{(Color online) Model of a quantum wire with length $L$, parabolic confinement and finite bulk region of width $\lbulk$.
The physical width $w(E_{\rm max})$ is indicated, where $E_{\rm max}$ is the largest energy considered.  }
\label{fig:wiremodel}
\end{figure}

As we seek to investigate the interplay and localization of edge states and bulk states in a quantum wire, we assume periodic boundary conditions in the $x$-direction along the wire  and choose 
\begin{equation}
V_{\mbox{\footnotesize conf}}(y)= \left\lbrace \begin{array}{cc}
 \frac{1}{2} m\omega_p^2 (y-\lbulk/2)^2& y\geq \lbulk/2 \\ 
 0& -\frac{1}{2}\lbulk<y<\frac{1}{2}\lbulk\\ 
 \frac{1}{2} m\omega_p^2 (y+\lbulk/2)^2& y\leq -\lbulk/2
\end{array} \right. 
\label{confinement}
\end{equation}
as confinement potential in the transversal direction (see Fig. \ref{fig:wiremodel}).
This model allows us to tune the confinement strength with the parameter $\omega_p$. The wire width is now defined by the bulk width $L_{\rm bulk}$. In the limit $L_{\rm bulk}=L$, we get the usual 2D model for the Quantum Hall effect \cite{Ando}, while for $L_{\rm bulk}=0$ we have the parabolic wire model  \cite{hajdubook}. 
In the limit of large confinement frequency   $ \omega_p > \omega_c$,  one approaches  hard-wall boundary conditions.
This type of confinement provides a smooth transition between the edge potential and the potential-free bulk region and renders the situation in real wires better than assuming hard wall boundaries. 

The physical width  $w$ of parabolic wires $w$ is a function of the Fermi energy $E$, 
\begin{equation}
w(E,\lbulk)=2\;\sqrt{ \frac{2E-\hbar\Omega}{\hbar\omega_B}}\;\frac{\omega_B}{\omega_p}\;l_B+\lbulk,
\label{physicalwidth}
\end{equation}
with $\Omega=\sqrt{\omega_p^2+\omega_B^2}$. It is obtained by finding the energy eigenvalue of the clean wire which is equal to $E$ and  has its  guiding center at $\pm w/2$.  
 We fix the basis width $L_{\rm basis}$ to be larger than the physical width $w(E_{\rm max})$ at 
 the highest considered  energy $E_{\rm max}$.   
 The total number of magnetic flux quanta in the model system is then fixed to $N_\phi=L_{\rm basis} L/(2\pi l_B^2)$.

\subsection{Wavefunction analysis}

 The Hamiltonian, Eq. (\ref{Hamiltonian_diag})
is diagonalized in the Landau representation with basis functions
\begin{equation}
\bra{\bf r}nX\rangle=\frac{1}{\left(l_B L\sqrt{\pi}2^n n!\right)^{1/2}}\mathrm{e}^{-\frac{(y-X)^2}{2l_B^2}} H_n\left(\frac{y-X}{l_B}\right)\;{\mathrm e}^{-\frac{iXx}{l_B^2}}
\end{equation}
Here we have assumed the Landau gauge for the vector potential.
 The matrix elements of the confinement  potential  in 
 the Landau representation are given in Appendix \ref{matrixelements}. 

The exact diagonalization of the Hamiltonian (\ref{Hamiltonian_diag}) yields eigenenergies $E_\alpha$ with corresponding wavefunctions
\begin{equation}
\psi_\alpha( {\bf r})=\sum\limits_{nX}\bra{\bf r}nX\rangle \langle nX\ket{\alpha}
\end{equation}
The spatial extention of these wavefunctions is characterized by their participation ratio
\begin{equation}
P_\alpha=\left (\lbulk L \int d^2r |\psi_\alpha( {\bf r})|^4\right )^{-1},
\end{equation}
which is small for localized states and large for extended states. Note that
in this definition,  $P_\alpha$ relates to the fixed bulk area $\lbulk L$ 
 while the  wave functions can cover a larger area due to the smooth confinement, so that 
 $P_\alpha > 1$ is possible for all states.

In clean 2D systems all states in a Landau level are degenerate. In a  disordered wire  
this degeneracy is lifted by the disorder, and at
the edges by the confinement potential. Therefore, 
localized  states in the tail of the Landau bands in the bulk region near the center of the wire coexist with  states at the edges at the same energy, and,  in principle,   mixing of states from the bulk with edge states is possible.


\begin{figure*}[htbp]
\hspace*{-1cm}\includegraphics[angle=-90, scale=0.6]{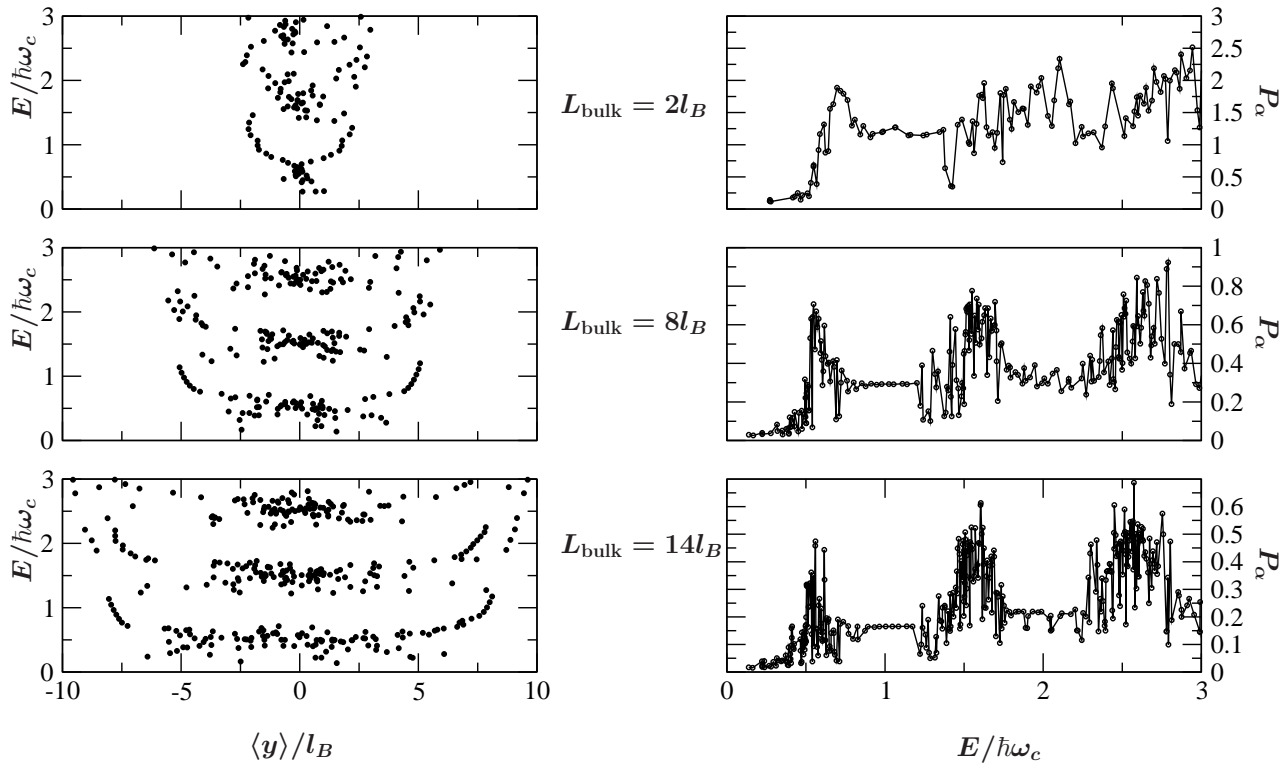}
\caption{Left: The energy eigenvalues $E_{\alpha}$  of all states  in  a wire of length $L = 40.1 l_B$ in a perpendicular magnetic field of 8 T for three different bulk widths are plotted versus the expectation value of the transversal position, $\langle \alpha \mid y \mid \alpha \rangle$. The disorder amplitude is fixed to $V_0=0.73\hbar \omega_c$ with $N_i = 150$ impurities   and the confinement energy is chosen to have the same 
magnitude with $\hbar\omega_p =0.73\hbar \omega_c$. The basis width is chosen as $L_{\mathrm{basis}}=1.5 ~  w(E_{\rm max})$, Eq.  (\ref{physicalwidth}),  with $E_{\rm max}=2\hbar\omega_c$.
 Right: Corresponding  participation ratio $P_\alpha$ versus $E_\alpha$.}
\label{fig:dispersionipn}
\end{figure*}
\begin{figure}[htbp]
\hspace*{-2cm}
{\includegraphics[angle=-90, scale=0.3]{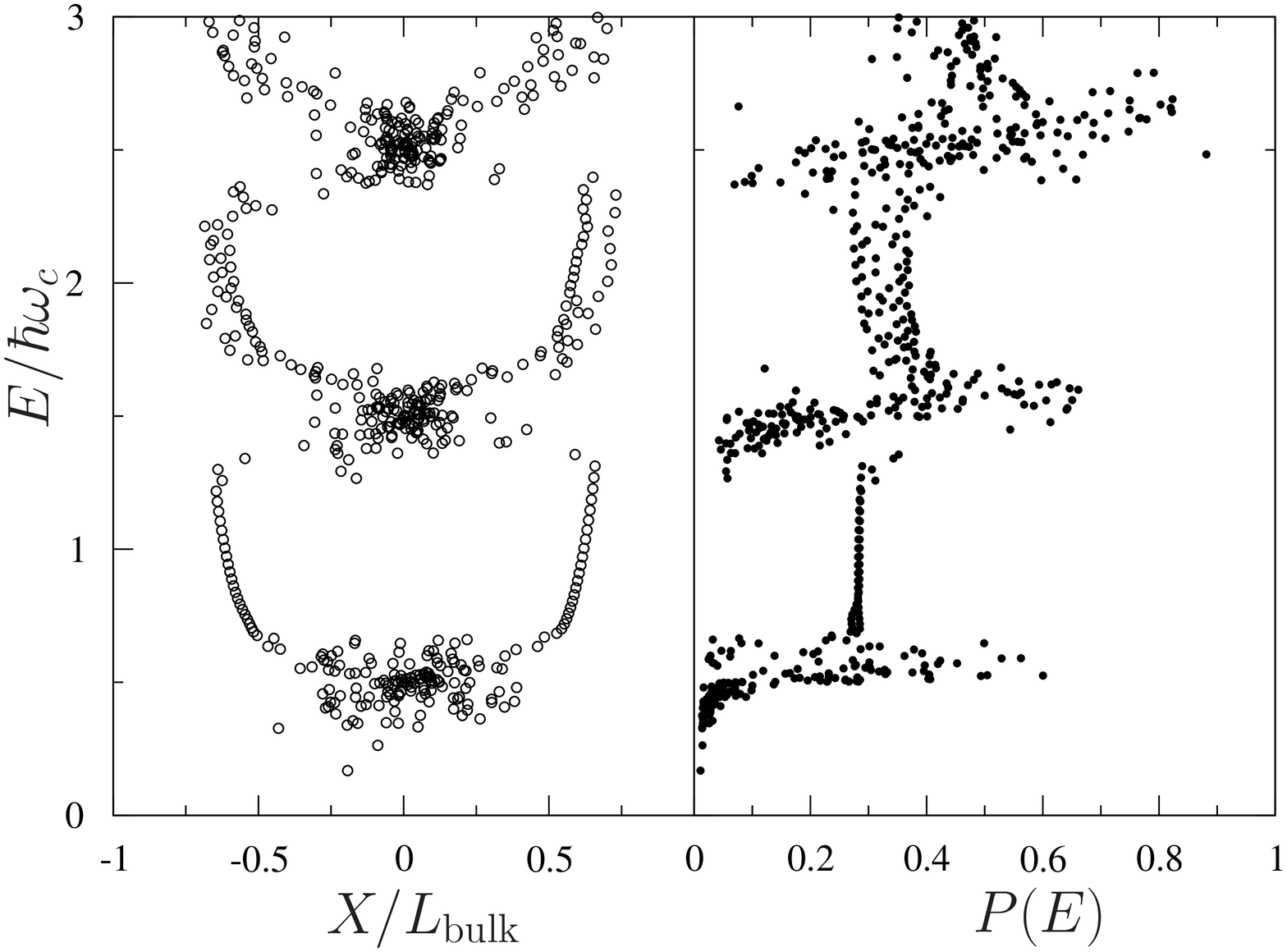}}
\caption{Energy dispersion $E(X)$ and corresponding participation ratio $P(E)$ for a longer system with $L=100 l_B$ and $\lbulk=8 l_B$ at $B=8$ T. The disorder is realized by 400 scatterers with uniformally distribted amplitude with maximal value,  $V_0=0.73\;\hbar\omega_c$, which equals the confinement energy $\hbar\omega_p=0.73\;\hbar\omega_c$.}
\label{fig:L100disp}
\end{figure}

 These features are clearly seen 
in the left part of Fig.  \ref{fig:dispersionipn}. For a 
wire of  length $L = 40.1 l_B$ in a perpendicular magnetic field of $B = 8 \mathrm{T}$, 
corresponding for $m^\ast = 0.067$ in units of the bare electron mass to $\hbar \omega_c = 13.82$ meV,  for three different bulk widths $L_{\rm bulk}$ the eigenenergies are  plotted versus the expectation value of their transversal position, $\langle \alpha \mid y \mid \alpha \rangle$. 
 Although we have chosen a smooth confinement potential, these results 
  are in good agreement with earlier results with short ranged
   disorder in Ref.  \onlinecite{ohtsuki}. 
  Obviously,  the edge states between the Landau bands are hardly  affected by the disorder potential. 
  There is  a coupling of edge states of the same chirality in the second and higher Landau bands which leads  to the formation of minibands in between  the Landau band, \cite{ohtsuki} 
  as seen most clearly in Fig. \ref{fig:L100disp}. There,   we show the same quantities for a
 longer system with $L=100 l_B$ and $\lbulk=8 l_B$ at $B=8$ T. The disorder is realized by 400 scatterers with   $V_0=0.73\;\hbar\omega_c$,  with the same value as the confinement energy $\hbar\omega_p=0.73\;\hbar\omega_c$.
  However, there is an abrupt shift of the center of the eigenstates towards the middle of the wire, when their energy is approaching  the middle of the Landau band. Still, one can not conclude, if this fact is mainly due to the backscattering between edge states from opposite edges, having opposite chirality, or if it is mainly due to a mixing with the bulk localized states. 

 In order to learn more about the nature of these states, we have calculated 
the Frmi energy dependence of the participation ratio for different bulk widths with fixed disorder potential and constant magnetic field as shown in the right part of Figs. \ref{fig:dispersionipn}, 
\ref{fig:L100disp}.
 It is observed  for all three widths  that the participation ratio  and  the eigenenergies, fluctuate as a result of disorder especially in the center of the wire. 
The participation ratio increases with energy in the  tails of the Landau bands and reaches a maximum close to 
the corresponding center energy (between $0.5\;\hbar\omega_c$ for $\lbulk\rightarrow\infty$ and $0.5\;\hbar(\omega_c^2+\omega_p^2)^{1/2}$ for $\lbulk\rightarrow 0$). The participation ratio 
 saturates to a constant value between the Landau bands, where only edges states exist, as  confirmed by comparison with the left side of Fig. \ref{fig:dispersionipn}. 

\begin{figure*}[htbp]
\hspace*{-2cm}
\includegraphics[angle=0, scale=0.8]{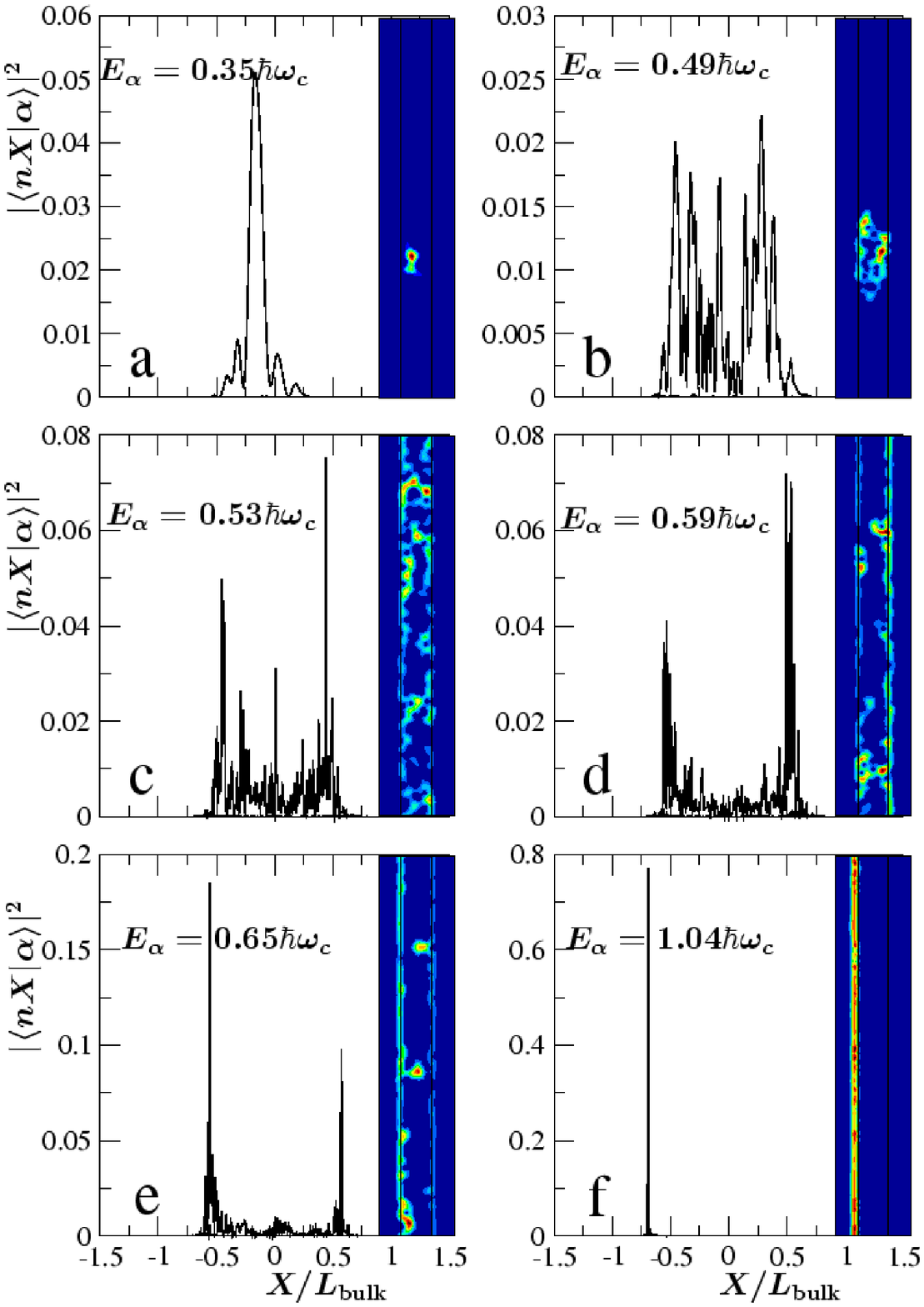}
\caption{(Color online) Basis state contributions to eigenstates at selected energies. The right insets show the correponding probability densities (blue for low, red for high values), solid lines mark the bulk region. System parameters are the same as in Fig. \ref{fig:L100disp}.}
\label{fig:evstat}
\end{figure*}

In the following, we scrutinize the localization behaviour in the different energy regions identified above by the energy dispersion  and the participation ratio.
To determine the nature of the states, we plot the basis state contributions and spatially resolved probabilities for a sample with $L=100 l_B$ and $\lbulk=8 l_B$ at typical energies, with disorder amplitude and confinement energy comparable to the cyclotron energy.
We concentrate on the lowest Landau level  and investigate wavefunctions at energies in characteristic regions of the participation ratio. The result is displayed in figure \ref{fig:evstat}. 

\begin{figure*}[htbp]
\subfigure[]{\includegraphics[angle=0, scale=0.6]{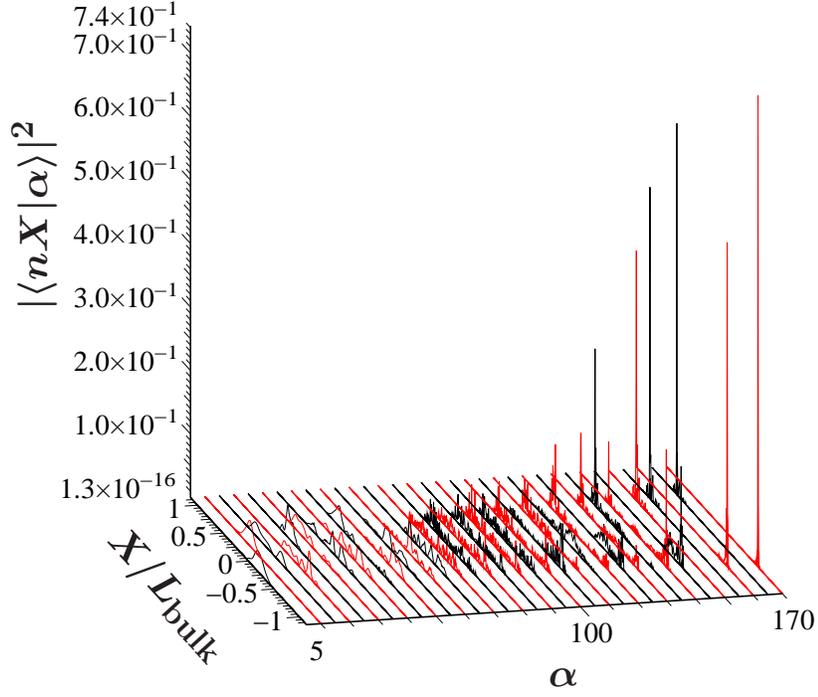}}
\subfigure[]{\includegraphics[angle=0, scale=0.6]{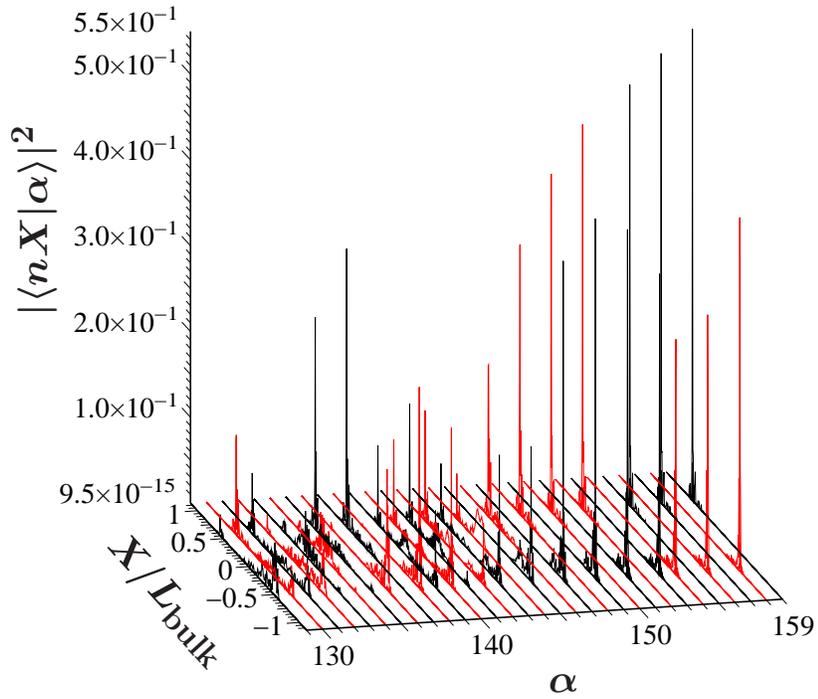}}
\caption{(Color online) Basis state contributions $|\bra{nX}\alpha\rangle|^2$ to eigenstates $\alpha$ at different energies (a) for every fifth state in the lowest Landau level, (b) for every state with energy between $0.6\;\hbar \omega_c$ and  $0.8\;\hbar \omega_c$. System parameters are the same as in Fig. \ref{fig:L100disp}. Different colors are used for adjacent curves in order to distinguish them clearly.}
\label{fig:evstat-all}
\end{figure*}

\begin{figure*}[htbp]
{\hspace*{-0.8cm}\includegraphics[angle=+90, scale=0.6]{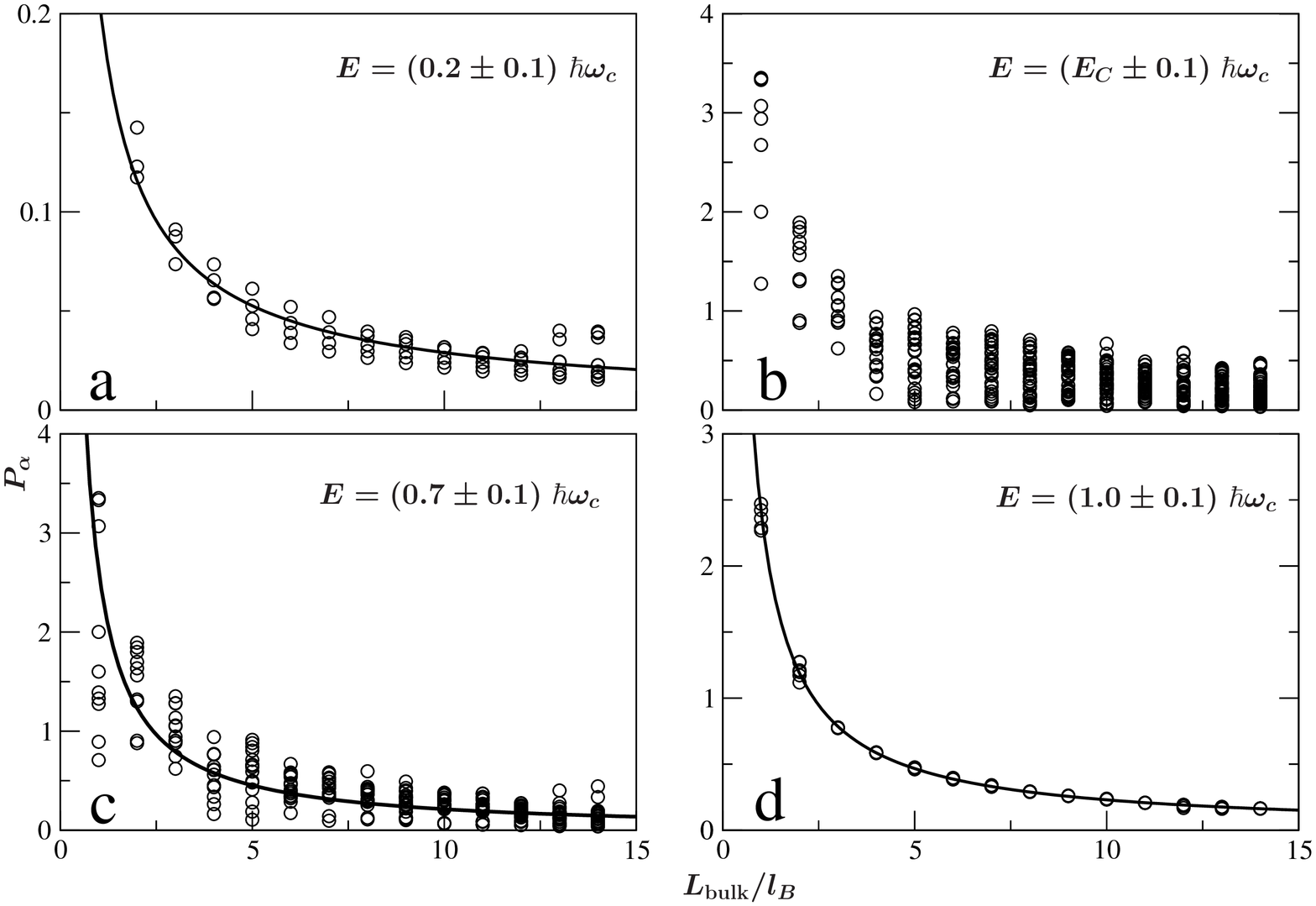}}
\caption{Participation ratio in dependence of bulk width for different energies. $E_C$ is determined for each bulk width as  the energy, at which the system reaches the maximal participation ratio in the lowest Landau band. System parameters are the same as in Fig. \ref{fig:dispersionipn}. Solid lines are fitting functions: (a) $P_\alpha=0.21\, (\lbulk/l_B)^{-0.86}$, (c) $P_\alpha=2.62\,(\lbulk/l_B)^{-1.09}$, (d) $P_\alpha=2.41\,(\lbulk/l_B)^{-1.02}$.}
\label{fig:ipnscale}
\end{figure*}

We find states at $E=0.3 \hbar \omega_c $ (Fig. \ref{fig:evstat}a) which are 2D localized, 
as confirmed by the fact that they have  contributions from basis states with guiding centers in the bulk region, only. In the band center around $E=0.5 \hbar \omega_c $ the participation ratio fluctuates strongly. In this region we find 2D localized states as well as  1D localized states with a localization length larger than the bulk width, but much smaller than the wire length (Fig  \ref{fig:evstat}b). In the latter case, basis states from bulk and edge region mix with comparable contributions. Furthermore, we can identify  states which cover the whole sample, as shown in Fig \ref{fig:evstat}c. These states couple to  all regions as well, although the contributions from the left and right edges seem to prevail slightly. The trend indicated by the maxima close to the edges  intensifies in the transition region (Fig. \ref{fig:evstat}d). At a specific energy, the contribution of the bulk states is small compared to the sharp maxima at the edges, while  the electron  is  found  with the same probability on the right \em or\  \rm the left edge of the wire  (Fig. \ref{fig:evstat}e). We believe that this nonchiral edge state is unique at least in the thermodynamic limit, of $w \rightarrow \infty$  and governs the new type of metal-insulator transition, the CMIT, 
 in quasi-1D quantum Hall wires. This state exhibits notable localization features: being an edge state concerning the 
 participation ratio, it has to be considered localized concerning the conductance, since current flows with equal probability, but 
 reversed sign on both edges. This behaviour is consistent with the sudden breakdown of the  conductance  observed in Fig. 
 \ref{phasenew}.

At higher energies, below the next Landau band,  edge states are formed as seen in Fig. \ref{fig:evstat}. These  states are found to be insensitive to  disorder.

This  sequence of transitions from 2D bulk states, quasi-1-D localized states, 
 states with peaks on both edges of the wire, and decoupled edge states is 
 visible  in   
 Fig. \ref{fig:evstat-all}a, where  we show the basis state contributions for every fifth state in the lowest Landau level. All features discussed above are seen clearly, with a remarkably narrow  transition region from  
 1D localized states to edge states, as one moves from state 125 to state 145.

In order to visualize this transition in detail, we plot in this interval  all states   in Fig. \ref{fig:evstat-all}b.
 The higher energy states are clearly edge states, which are decoupled from the bulk, 
 and are altenatingly  located either 
on the left or on the right side of the wire. 
 The lower the energy, the smaller becomes that peak in  intensity at the edge.
  Still,  each 
  state stays  located in the edge region, 
  with only a small coupling to  the nearest part of the bulk.   
Then, suddenly, at state  ($\alpha=143$ in Fig.  \ref{fig:evstat-all}), there appear
 two  peaks of comparable amplitude 
 on both edges, while the contribution of the bulk is still small. All states $\alpha=139-143$ share the two pronounced peak at the edges, while the 
 bulk contribution increaes only slowly with lowering the energy. 
 Before the  transition to quasi-1D-states with more or less uniform distribution across the bulk, there is a reappearance of edge  like states 
 $\alpha=135-138$, which we attribute to mesoscopic fluctuations due to the 
  random distribution of disorder in this rather  mesoscopic sample. 
The transition which we observed here  happens thus rather smooth 
 as compared to the  sharp transitions 
  in the transfer matrix results shown in Fig. \ref{phasenew}. 
  This  can be attributed to the fact 
  that the finite system with $ L = 100 l_B$ which we have diagonalized here is 
  much smaller than the system 
  which was handled by the transfer matrix method.
   As expected 
   far away from the thermodynamic limit, $L,w \rightarrow \infty, L/w={\rm const.}$, the transition 
   occurs in a finite energy interval rather than  at a single point.   
   Finite size effects can be revealed further   by 
 studying  the dependence of the states on the bulk width. 
 
 To this end  we next study   the system size dependence of the participation ratio.
 Fig. \ref{fig:ipnscale} shows the  participation ratio of all states in a given energy interval for systems with different bulk widths $\lbulk$. Disorder configuration, wire length and confinement energy $\hbar\omega_p$  are kept fixed for all the systems.
 As a characteristic example for the behaviour in the low energy region, we investigate states in an interval around  energy  $E=0.2 
 \hbar\omega_c$ (Fig. \ref{fig:ipnscale}a). In this region, the participation ratio scales with the wire width approximately as 
 $P\propto\lbulk^{-1}$. This is in agreement with the expected scaling of 2D localized states which give a contribution 
 $\psi_\alpha^2 \propto 1/\xi_{2D}^2$, only within a localization area  $\xi_{2D}^2$
  where $\xi_{2D}$ is the 2D localization length of the wavefunction, which is 
  independent of 
 the  wire length $L $ and width $ \lbulk$.  It follows that, 
  \begin{equation}
  P_{2D} \propto  \xi_{2D}^2 L^{-1}\lbulk^{-1} \ll 1,
  \end{equation}
  in good agreement with Fig. \ref{fig:ipnscale}a.
 
 The behaviour changes in the center of the Landau band
  (Fig. \ref{fig:ipnscale}b). There, the  
 density of states is higher, and the disorder results in   a   wide range
 of participation 
 ratios. For large bulk widths, $\lbulk>5 l_B$, the range of participation 
 ratios becomes  constant and saturates to a finite value.
  Note that   quasi-1-D localized states cover 
   approximately an area $\lbulk \xi_{1D} \sim g \lbulk^2$, 
   and  contribute in this area with probability density 
 $\psi_\alpha^2 \sim 1/(\xi_{1D} \lbulk)$.
  As a result, one expects for quasi-1-D localized states, according to Eq. (\ref{1db}),
  \begin{equation}
  P_{1D} \propto  \frac{\xi_{1 D}}{L}  \sim  g \frac{\lbulk}{L},
  \end{equation}
  increasing linearly with $\lbulk$. 
   When the wire is comparable or shorter than the quasi-1D localization length, 
    however, the participation rato shows rather the behavior of 2D extended states
     which cover the whole wire area $L w$ with probability density $1/(L w)$,
      yielding the typical participation rato of extended states 
      \begin{equation}
      P_{ext} \sim w/\lbulk 
  = \mathrm{const} > 1,
  \end{equation}
   being independent of the  width  $\lbulk $. Note that for extended statesone would expect $P_\alpha=w/\lbulk$, which according
   to equation (\ref{physicalwidth}) converges to 1 for $\lbulk\rightarrow\infty$.
   Whereas $P_{\alpha}$ in Fig. \ref{fig:ipnscale}b is indeed seen to saturate to a constant 
    mean value, this value is found not to exceed 1. This is consistent with the fact that the wavefunction is multifractal.\cite{klesse}
  
 The scaling of the participation ratios  in the high energy tail of the lowest Landau band (Fig. \ref{fig:ipnscale}c,d) is again a power 
 law $P\propto\lbulk^{-1}$, but with an absolute value much larger than in Fig. \ref{fig:ipnscale}a. This resembles the expected 
 feature for edge states, which cover an area $l_B L$ with probability 
 density $1/(l_B L)$, yiedling, 
 \begin{equation}
P_{\rm edge} \propto l_B/\lbulk,
 \end{equation}
 which is both in magnitude and in the functional dependence 
  on $\lbulk$ in good agreement with Fig. \ref{fig:ipnscale}d.
  Note that in the transition region, Fig. \ref{fig:ipnscale}c, 
   the large mesoscopic fluctuations do not allow to distinguish 
    characteristic features of the nonchiral edge states at the transition, 
     the functional dependence on $\lbulk$ is that expected for edge states
      and localized bulk states alike, which both coexist in this energy region, 
       as we had seen above in Fig. \ref{fig:evstat-all}.
  
In summary, our model allows to study the mutual influence 
  between the states in the bulk region,  where the influence of  the disorder potential
   is strong, and states in the edge region, where the confinement potential prevails. We found that in the  narrow energy region of the CMIT the disorder-induced coupling between the edges creates nonchiral edge states which have comparable  weights
    on both edges, but   only a negligible  mixing with the  bulk.





\section{Conclusions}
We conclude, that in quantum Hall bars of finite width $w \ll \xi_n$ at low
temperatures  quantum phase transitions occur between
 extended chiral  edge states and a quasi-1D insulator. 
These are driven by the  crossover from 2D to 1D localization of  bulk states. These
metal-insulator transitions
{\it resemble } first-order
phase transitions in the sense that the localization length
 abruptly jumps between exponentially large  and finite values, 
  which we confirmed by calculating the edge state localization length, explicitly.
 In the thermodynamic limit, {\it fixing the aspect ratio $c = L/w$, 
 when sending $L \rightarrow \infty$, then $c \rightarrow \infty$, 
  the two--terminal  conductance 
 jumps between exactly integer values and     zero}. 
 The transitions occur at energies where the localization
length of  bulk states is equal to the geometrical wire width. Then,  $m$ edge
states  mix and  electrons are free to diffuse between
the wire boundaries but become Anderson localized along the wire.
 Close to that  transition we found with exact diagonalisation studies  that  
  particular states  exist, 
  which are superpositions of edge states with opposite chirality, 
   with   an order of magnitude 
    smaller  bulk contribution. Although this state
    is located at the edges, it is a nonlocal state, having comparable weights 
     on opposite sides of the sample. Thus, it can have a mesoscopic extension 
      across the width of the Hall bar, if it is more narrow 
       than the phase coherence length. 
The Chiral Metal--Insulator Transition is of mesoscopic nature since, 
 at finite temperature, the
phenomenon of the CMIT  can only be observed, when the phase coherence length exceeds the
quasi-1D localization length in  centers of  Landau bands, $L_{\varphi}
> \xi_n$. 
 One then should observe 
transitions of the two-terminal resistance from integer quantized plateaus,
$R_n = h/n e^2$ to a Mott variable-range hopping regime of exponentially
diverging resistance. Such experiments would yield information about the coupling between edge and bulk 
states in quantum Hall bars.
At higher temperature, when $L_{\varphi}< \xi_n$, the conventional form of the
integer quantum Hall effect is recovered  \cite{klitzing}.
   
  We have studied the modification of the CMIT by correlations in the
disorder
 potential and due to interactions. These results will be
 presented in a subsequent publication.
 
{\bf Acknowledgments}
 We acknowledge useful discussions with M. E. Raikh,
  A. MacKinnon, and B. Huckestein.  
 This research was
supported by  German Research Council (DFG), Grant No. Kr 627/10,  
 Schwerpunkt "Quanten-Hall-Effekt",
 and  by EU TMR-network Grant. No.  HPRN-CT2000-0144.

 \appendix
 
\section{Matrix elements for confinement in Landau representation}
\label{matrixelements}


The matrix element of the confinement potential defined by equation (\ref{confinement}) in Landau representation $\bra{nX}V_{\mbox{\footnotesize conf}}\ket{n'X'}$ is given by
\begin{widetext}
 \begin{equation}
 \bra{nX}V_{\mbox{\footnotesize conf}}\ket{n'X'}=\delta_{XX'}\frac{1}{\sqrt{\pi}l_B 2^{(n+n')}n!n'!} \left [M_{nn'}\left 
 (\frac{\lbulk}{2},X\right )+(-1)^{(n+n')}M_{nn'}\left (\frac{\lbulk}{2},-X\right )\right ],
 \end{equation}
 
 and
 \begin{eqnarray}
 M_{nn'}\left (\frac{\lbulk}{2},X\right)&=&\int\limits_{\lbulk/2}^{\infty} dy\,\,\mathrm{e}^{-\frac{(y-X)^2}{l_B^2}}H_n\left 
 (\frac{y-X}{l_B}\right )H_{n'}\left (\frac{y-X}{l_B}\right ) \left (\frac{y-\lbulk/2}{l_B}\right)^2\\
 &=& l_B \int\limits_{b}^{\infty} d\xi\,\,\mathrm{e}^{-\xi^2}(\xi-b)^2 H_n(\xi) H_{n'}(\xi)\;\; ,
 \label{confmatrix2}
 \end{eqnarray}
 \end{widetext}
where $\xi=(y-X)/l_B$ and $b=(\lbulk/2-X)/l_B$.
 
By expanding all polynomials in Eq.  (\ref{confmatrix2}) in monomials in $\xi$ using the relation
\begin{equation}
H_n(\xi)=n!\sum\limits_{m=0}^{[\frac{n}{2}]} (-1)^m \frac{2^{n-2m}}{m!(n-2m)!}\xi^{n-2m},
\end{equation}
where $[x]$ denotes the largest integer smaller than $x$, one gets
\begin{widetext}
\begin{eqnarray}
M_{nn'}(b,X)&=& l_B\,\, n!n'!\sum\limits_{l=0}^{[\frac{n}{2}]}\sum\limits_{k=0}^{[\frac{n'}{2}]}\Biggl[
(-1)^{l+k}\frac{2^{n-2l+n'-2k}}{l!k!(n-2l)!(n'-2k)!} \nonumber\\
 &\times&\left (f^{(2+n-2l+n'-2k)}(b)-2bf^{(1+n-2l+n'-2k)}(b)+b^2 f^{(n-2l+n'-2k)}(b)\right ) \Biggr ]\nonumber
\end{eqnarray}
\end{widetext}
In the last expression,
\begin{widetext}
\begin{equation}
f^{(M)}(b)=\int\limits_{b}^{\infty} d\xi\,\,\xi^M \mathrm{e}^{-\xi^2}= \frac{M-1}{2}f^{(M-2)}(b)+b^{M-1}\frac{1}{2}\mathrm{e}^{-b^2}
\end{equation}
\end{widetext}
This recursive formula can be obtained by repeated partial integration and is valid for even and odd $M>1$. An explicit evaluation requires the inital expressions 
\begin{eqnarray}
f^{(0)}(b) &=& \frac{1}{2}\mathrm{e}^{-b^2}\\
f^{(1)}(b) &=& \frac{1}{2}\sqrt{\pi}\;\mbox{\rm erfc} (b)\,\, ,
\end{eqnarray}
with the complementary error function
\begin{equation}
\mbox{\rm erfc} (b)=\int\limits_{b}^{\infty}d\xi\,\, \mathrm{e}^{-\xi^2}.
\end{equation}

\end{document}